\def\sii{S~{\sc ii}}

\def\hi{H~{\sc i}}

\def\feii{Fe~{\sc ii}}

\def\siii{Si~{\sc ii}}

\def\znii{Zn~{\sc ii}}

\documentclass[useAMS,usenatbib,epsfig]{mn2e}

\usepackage{rotating}
\usepackage{float}
\usepackage{natbib}
\usepackage{verbatim}
\usepackage{hyperref,graphicx}

\title[Silicon and Iron dust in GRBs]{Silicon and iron dust in gamma-ray burst host galaxy absorbers}
\author[T. Zafar et al.] {T. Zafar$^{1}$\thanks{e-mail:tayyaba.zafar@mq.edu.au}, K. E. Heintz$^{2}$, A. Karakas$^3$, J. Lattanzio$^3$, A. Ahmad$^1$ \\
$^1$ Australian Astronomical Optics, Macquarie University, 105 Delhi Road, North Ryde, NSW 2113, Australia \\
$^2$ Centre for Astrophysics and Cosmology, Science Institute, University of Iceland, Dunhagi 5, 107 Reykjav\'ik, Iceland \\
$^3$ Monash Centre for Astrophysics, School of Physics and Astronomy, 10 College Walk, Monash University 3800, Australia
}

\begin{document}


\pagerange{\pageref{firstpage}--\pageref{lastpage}} \pubyear{2019}
\maketitle
\label{firstpage}
\begin{abstract}
Depletion studies provide a way to understand the chemical composition of interstellar dust grains. We here examine 23 gamma-ray bursts (GRB) optical afterglow spectra (spanning $0.6\leq z \leq5.0$) and compare their silicon and iron dust-phase column densities with different extinction curve parameters to study the composition of the interstellar dust grains in these high-redshift GRB host galaxies. The majority of our sample (87\%) show featureless extinction curves and only vary in shape. We observe strong correlations (with $>96\%$ significance) between the total-to-selective extinction, $R_V$, and the dust-phase column densities of Si and Fe. Since a large fraction of interstellar iron is locked in silicate grains, this indicates that high Si and Fe depletion leads to an increase in the fraction of large silicate grains and vice versa. This suggests that silicates play a vital role to induce the entire extinction at any wavelength. On the other hand, the far-UV extinction is usually attributed to the presence of small silicates. However, we find no trend between the far-UV parameter of the extinction curve, $c_4$, and the abundance of Si and Fe in the dust phase. We, therefore, propose that the far-UV extinction could be a combined effect of small (probably nanoparticles) dust grains from various species.
\end{abstract}

\section{Introduction}
Dust is an essential component of the interstellar medium (ISM) of galaxies \citep{dorschner95} and is strongly linked with star formation \citep[e.g.,][]{cortese12,corre18}. It is therefore important to understand the chemical composition of interstellar dust grains. The interstellar dust grains are primarily made up of O, C, Si, Mg, and Fe elements \citep{decia16} which are introduced into the ISM during stellar evolution via stellar winds or at the end of star's life. Depletion studies of highly abundant species, i.e. silicon and iron, provide direct evidence that refractory elements go in and out of the dust-phase in the ISM \citep{jenkins87,jenkins09}. These refractory elements are introduced into the ISM by stars during their stellar evolution or at the end of the life of a star and then completely condense into a solid form. Particularly silicon makes silicates by combining with different elements and may comprise up to 70\% of the total dust grains mass in the local ISM \citep{weingartner01}. Silicate and carbonaceous grains then evolve through many cycles of accretions, shattering, coagulation, and erosion happening in the ISM.

The Milky Way, Large and Small Magellanic Clouds (LMC and SMC) exhibit distinct extinction curves \citep{ccm89,pei92,gordon03,valencic03,sofia05,fm07}, showing large variations in the strength of the 2175\,\AA\ extinction bump and the slope of the induced extinction curve. \citet{stecher65} first discovered the 2175\,\AA\ extinction bump in the extinction curves of the Milky Way. The feature becomes weaker in the LMC and SMC display a featureless extinction curve with a steep rise into the UV. The 2175\,\AA\ bump has been attributed to absorption by carbonaceous dust grains produced by star formation \citep{draine03}. The slope of the extinction curve is characterised by the total-to-selective extinction parameter, $R_V$. In the Milky Way, $R_V$ parameter ranges from $\sim2-5$ \citep{fm07}, where a small value of $R_V$ corresponds to a steep extinction curve. Large average dust grains therefore produce a \lq flatter\rq\ extinction curve, resulting in large values of $R_V$ and vice versa \citep{valencic04}.

Gamma-ray bursts (GRBs) are powerful probes to study the ISM of high-redshift galaxies \citep[e.g.,][]{prochaska07,postigo12}. At the same time, because of their simple power-law spectra, GRBs provide a unique tool to study the absolute extinction curves of their environments \citep[e.g.][]{kann06,greiner11,schady12,zafar11,covino13,zafar18a}. Comparing extinction curve parameters with depletion of various refractory elements can provide us clues to the composition and distribution of interstellar dust grains in GRB host galaxy environments. For example, it has previously been found that the strength of the 2175\,\AA\ extinction bump observed in GRB and quasar absorbers is correlated with the amount of neutral carbon \citep{kruhler08,ardis,perley11,zafar12,ledoux15,ma18,heintz19a}. Recently, \citet{heintz19} also found an additional positive trend between the bump strength and $R_V$, so in combination with the relation to the amount of neutral carbon this suggests that the 2175\,\AA\ bump is produced by large dust grains and molecules in the cold and molecular gas-phase of the ISM.

From the observations of 16 Galactic sightlines, \citet{haris16} claimed that that the silicon depletion correlates with the 2175\,\AA\ bump and the far-UV extinction rise. However, more recently \citet{mishra17} found that the bump strength correlates with carbon abundance instead of silicon and there is no correlation of far-UV extinction with either silicon or carbon abundance using a silicate-graphite model on a larger sample. This discrepancy could be related to the different methods adopted for determining the silicon dust depletion and usage of different interstellar abundances of silicon in the two studies. The depletion of various elements and its connection to dust and metals has also been studied previously in GRBs and quasar absorbers \citep[e.g.,][]{savaglio04,vladilo06,vladilo11,decia13,zafar13,decia16,decia18,zafar19d}. For GRBs, \citet{heintz19} already compared carbon against bump strength and $R_V$. GRBs usually exhibit featureless extinction curves \citep[e.g.][]{kann06,greiner11,schady12,covino13,zafar18a}. We here aim to compare for the first time silicon and iron depletion against different extinction curve parameters to study the chemical composition of dust in the average GRB environment further.

We carefully selected a sample of GRB afterglows with measurements of extinction curve parameters using the parametric dust model of \citet{fm90} and with available silicon and/or iron abundances. In \S2 we present our sample and methods used for investigation. In \S3 we provide the results together with our discussion. Conclusions are given in \S5. Throughout the paper, given errors are 1$\sigma$ unless stated otherwise.

\section{Sample Selection and method}

For consistency, we included GRBs in our sample that have been examined using the same dust model to fit their spectral energy distributions (SEDs). Also for these GRBs, we selected the studies where the GRB optical extinction is derived from simultaneous SED fitting to X-ray$-$to$-$optical/NIR data using either a single or broken power-law. This way, the intrinsic slope of the GRB is constrained by the X-ray data, providing a reliable method to determine the extinction curve of the burst. We, therefore, used GRBs from \citet{zafar18a,zafar18b} and \citet{heintz19} using a freeform parametric dust model of \citet{fm90}. See \citet{zafar18a} for more details on the extinction model. In the following we only consider the SED-derived extinction curve parameters $R_V$, $c_4$ (far-UV parameter), and $A_V$ for the GRBs in our sample. We first searched in the literature for the refractory element (silicon: \siii\ and/or iron: \feii) abundances of these GRBs. We found that only 16 GRBs have published abundances of these refractory elements to enter into our sample. In addition, we included 7 more GRBs with extinction curve measurements from Zafar et al. (in prep) with available measurements of Si and/or Fe abundances. A consistent method to \citet{zafar18a} is used in Zafar et al. (in prep), the latter including the entire X-shooter GRB afterglow legacy sample presented in \citet{selsing19}. This make up a total of 23 GRBs in our sample within the redshift range of $0.6\leq z \leq5.0$ (see Table \ref{GRB}). 

We then looked for volatile element abundances (sulphur: \sii\ and/or zinc: \znii) for these 23 cases to make pairs with refractory element abundances to derive depletion (i.e. $\delta_X={\rm log} (N(X)/N(Y))-{\rm log} (X/Y)_\odot$; where $X=$ refractory and $Y=$ volatile element). However, we find only 8 cases with sulphur and 17 cases with zinc abundances measurements.

We derived dust-phase column densities (i.e. $N(X)_{\rm dust}=(1-10^{\delta_X})N(Y)(X/Y)_\odot$) of Si and Fe for our GRBs where abundance pairs are available using the depletions. For the remaining GRBs without abundance pairs, we instead used the tight dust-to-metals correlation of \citet{zafar19d}. This relation was derived by using the observations of 93 GRB afterglows and quasar absorbers ($z\approx0.4-6.3$). The GRBs included in their sample are in the range of $z\sim1.1-6.3$ where $A_V$ is derived from the afterglow SED fitting and metal abundances are obtained from absorption line analysis. Using the \citet{zafar19d} relation, we first determine the `total' column density of refractory elements. This is done by using $A_V$ values for each case and the estimated correlation of $A_V/N($\hi$)\times10^{[\rm{Si}_{tot},\rm{Fe}_{tot}/H]}=(4.91\pm0.98)\times10^{-22}$. The dust-phase column density of Si and Fe is then calculated by subtracting the observed column density of refractory elements from the total (since $N$(X)$_{\rm tot} =$ $N$(X)$_{\rm dust} + N$(X)$_{\rm obs}$). The above mentioned correlation, instead of the direct relation of $A_V$ and dust-phase column density of the refractory elements, is used as it is the strongest relation, matches to the Local Group relation, and provides an opportunity of using the observed refractory element abundances. 

The majority of the spectra in our sample have been obtained with the VLT/X-shooter with medium spectral resolution where {\it hidden} saturation could be a concern and a narrow 3$-$5\,km\,s$^{-1}$ component may not be resolved. The column densities are derived by fitting Voigt profiles to multiple transitions of the ions simultaneously where weak lines provide the accurate measurements of column densities. The studies from where abundances are included in our sample have paid attention while deriving abundance measurements. A narrow hidden component could in principle underestimate the element column density by $\sim$0.1\,dex \citep{zafar11c,rafelski12,cucchiara15,wiseman17} and hence overestimate the dust-phase column density. Our error bars on the dust-phase column densities are larger than 0.1\,dex so this issue will not affect our results significantly. Moreover, our chosen volatile elements S and Zn are also mildly depleted, particularly, in the presence of high molecular fraction. We refer the reader to the discussions in \citet{decia16} and \citet{zafar19d}. We have only 4 cases (GRB\,120815A, GRB\,120909A, GRB\,121024A, and GRB\,141109A) reported with the presence of molecules \citep[see][]{bolmer19} and for these the dust-phase abundances of Si and Fe may be slightly underestimated.

The relation between the visual extinction. $A_V$, and dust-phase column density of refractory elements has been studied previously for quasar absorbers \citep{vladilo06} and GRB systems \citep{decia13}. We here compared our dust-phase abundances of Si and Fe against other extinction parameters, namely $1/R_V$ and the far-UV component ($c_{4}^{\prime}=c_4/R_V)$. We consider $c_{4}^{\prime}$ here because in the parametrization of \citet{fm90} all the $c$ parameters are normalised by $R_V$. Therefore, $c_{4}^{\prime}$ better describe the shape of the entire far-UV component.

\begin{figure}
\begin{center}
{\includegraphics[width=\columnwidth,clip=]{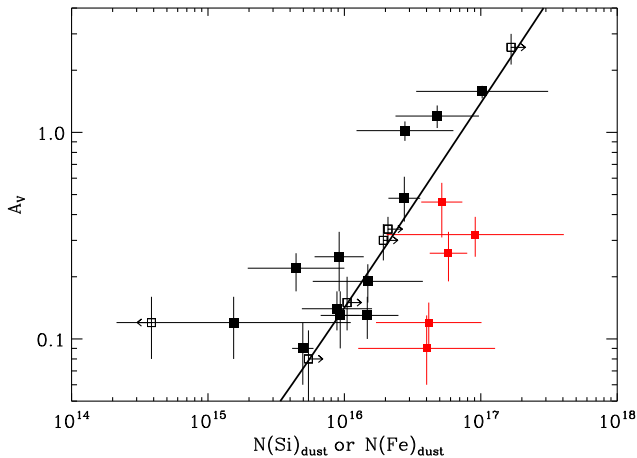}}
\caption{Extinctions as a function of Si (black) and Fe (red) dust-phase column densities for GRBs. The $N$(Fe)$_{\rm dust}$ is only considered for the cases where there is no $N$(Si)$_{\rm dust}$ measurement available. The solid line represents the Pearson linear correlation with a best-fit slope of $y/x=(1.39\pm1.03)\times10^{-17}$. The filled squares are measurements while the open squares are limits.}
\label{avsidust}
\end{center}
\end{figure}

\section{Results and Discussions}
Fig. \ref{avsidust} shows that for our sample, extinction is linearly correlated with the dust-phase column density of silicon or iron. This trend is already known for GRBs and quasar  absorbers \citep{vladilo06,zafar19d}. We fit a Pearson linear relation to our data and find that the data fit well (significance $\alpha>99\%$ and correlation coefficient $\rho=+0.83$) with a slope of $y/x=(1.39\pm1.03)\times10^{-17}$. This is consistent with the relation found by \citet{zafar19d} and expected as some of the dust phase column densities are computed using the $A_V$ values and 13 out of 23 GRBs are included in both studies. This suggests that, overall, the dust grains producing the observed extinction are linked to the abundance of Si and Fe in the dust phase.

\begin{figure*}
\begin{center}
{\includegraphics[width=0.8\textwidth]{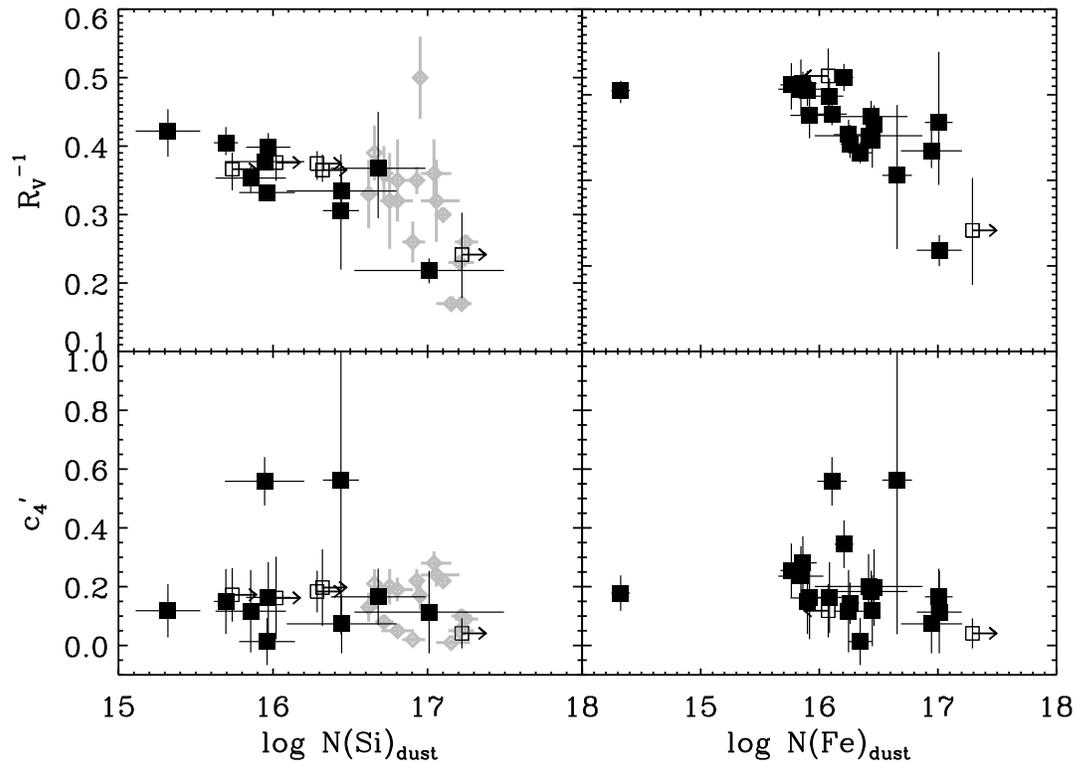}}
\caption{Inverse total-to-selective extinction ($R_V^{-1}$; {\it top panels}) and far-UV component ($c_{4}^{\prime}$; {\it bottom panels}) compared against dust-phase column densities of silicon and iron for GRB afterglows. The filled squares represent measurements and open squares indicate limits. The Galactic silicon dust-phase column densities from \citet{haris16} are shown as gray diamonds for a comparison.}
\label{comp}
\end{center}
\end{figure*}
 
\subsection{$R_V$ against dust-phase Si and Fe}\label{rvdust}
We compared dust-phase column densities of Si and Fe for GRBs with $R_V^{-1}$ and far-UV extinction component, $c_{4}^{\prime}$ (see Fig. \ref{comp}). As the top panels of Fig. \ref{comp} show, $R_V^{-1}$ anti-correlates with both $N$(Si)$_{\rm dust}$ and $N$(Fe)$_{\rm dust}$. A linear regression analysis of $R_V^{-1}$ versus $N$(Si)$_{\rm dust}$ and $N$(Fe)$_{\rm dust}$ for the GRBs in our sample yields a Pearson's correlation coefficient $\rho=-0.63$ (significance $\alpha\approx96\%$) and $\rho=-0.48$ (significance $\alpha\approx96\%$), respectively. Since $R_V$ is the absolute extinction to color excess ratio ($R_V=A_V/E(B-V)$), and it correlates with the shape of extinction curves \citep{ccm89}, this relation suggests that the Si and Fe dust components correlate well with the extinction over the entire wavelength range from the visible to the far-UV. The value of $R_V$ is directly related to the size of the dust grains: Large $R_V$ values are produced by larger grains in the environment and small $R_V$ values are characterised by smaller grains. Low $R_V$ sightlines produce very steep extinction curves in the UV/optical and comprise a large fraction of small dust grains and vice versa \citep{valencic04}. Our results suggest that with an increase in Si and Fe depletion, the average grain sizes will also increase in the ISM. This is consistent with most of the Si and Fe being depleted from the gas-phase and accreting onto dust grains. In particular, \citet{draine90} proposed that in dense environments with high depletion, small grains coagulate onto large grains. It is also worth noting that shattering during grain-grain collisions in shock waves can return dust grain mass from large to small grains \citep{weingartner01}.

Previously, \citet{haris16} reported a weak trend (with a Pearson's correlation coefficient $\rho=-0.13$ and significance $\alpha\approx37\%$) between $R_V$ and Si depletion. We plotted their data in Fig. \ref{comp} in grey colour, however, the \citet{haris16} data are not included in our correlation analysis. We attempted to compare their results with our work. We used their Tables\,$3-4$ to compare $R_V$ with the dust-phase column density of Si for the Galactic sightlines. This is done to make their work comparable to this study. Moreover, extinction is generated by dust particles and instead of comparing relative measurement of refractory element in dust (depletion) it is useful to compare column densities of their atoms in the dust. Their log $N$(Si)$_{\rm dust}$ ranges from $\sim16.5-17.3$ and $R_V$ ranges from $\sim2-6$. A linear regression analysis of their data yields a Pearson's correlation coefficient $\rho=-0.57$ (significance $\alpha\approx98\%$), consistent with our findings. Top-left panel of Fig. \ref{comp} also shows that the trend is more apparent at log $N$(Si)$_{\rm dust}>16$ as seen in the data of \citet{haris16}. Our relation between $R_V^{-1}$ and silicon and iron in dust  suggests that dust grain size varies as a function of the amount of Si and Fe in the dust-phase.


\begin{table*}
\caption{GRB afterglow data used to estimate N(Si, Fe)$_{\rm dust}$. The columns are given as: (1) GRB name, (2) extinction ($A_V$), (3) redshift, (4-7) observed \siii , \feii , \sii , \znii\ column densities, (8) far-UV parameter $c_4$, (9) total-to-selective extinction $R_V$, and (10) References (where the first reference is for dust extinction measurement and second reference(s) is for metal column density measurement).}     
\label{GRB} 
\centering     
\setlength{\tabcolsep}{5pt}
\renewcommand{\footnoterule}{}  
\begin{tabular}{l c c c c c c c c c c}  
\hline\hline                        
GRB & $A_V$ & $z$ & log N(\siii) & log N(\feii) & log N(\sii) & log N(\znii) & $c_4$ & $R_V$ & Refs.\\
 &  mag  & & cm$^{-2}$ & cm$^{-2}$ & cm$^{-2}$ & cm$^{-2}$ & \\
\hline
061121 & $0.46^{+0.11}_{-0.15}$ & 1.315 & $\cdots$ & $16.20\pm0.03$ & $\cdots$ & $13.76\pm0.06$ & $0.34\pm0.12$ & $2.88^{+0.23}_{-0.27}$ & 1\\
070802 & $1.20\pm0.15$ & 2.455  & $16.60\pm0.30$ & $16.10\pm0.10$ & $\cdots$ & $13.60\pm0.60$ & $0.45\pm0.09$ & $2.72^{+0.61}_{-0.54}$ & 1, 2 \\
080605 & $0.48^{+0.13}_{-0.11}$ & 1.640 & $15.88\pm0.10$ & $14.66\pm0.11$ & $\cdots$ & $13.53\pm0.14$ & $1.84\pm0.50$ & $3.27^{+0.88}_{-0.92}$ & 1, 3 \\
080607 & $2.58^{+0.42}_{-0.45}$ & 3.037 & $>16.34$ & $>16.70$ & $\cdots$ & $\cdots$ & $0.17\pm0.05$ & $4.14^{+1.05}_{-1.09}$ & 1, 3 \\
090313 & $0.30\pm0.06$ & 3.374 & $>15.42$ & $15.08\pm0.30$ & $\cdots$ & $\cdots$ & $0.49\pm0.07$ & $2.67^{+0.13}_{-0.17}$ & 4, 5 \\
090809 & $0.12\pm0.04$ & 2.737 & $16.15\pm0.07$ & $15.75\pm0.07$ & $15.85\pm0.60$ & $13.25\pm0.11$ & $0.62\pm0.09$ & $2.43^{+0.15}_{-0.17}$ & 3, 6\\
100219A & $0.14\pm0.03$ & 4.668 & $15.15\pm0.25$ & $14.73\pm0.11$ & $14.25\pm0.15$ & $\cdots$ & $1.48\pm0.08$ & $2.65\pm0.09$ & 3, 4 \\
100418A & $0.12\pm0.03$ & 0.624 & $\cdots$ & $15.62\pm0.10$ & $\cdots$ & $13.59\pm0.25$ & $0.68\pm0.09$ & $2.42^{+0.08}_{-0.10}$ & 4, 7 \\
100901A & $0.25\pm0.08$ & 1.408 & $15.96\pm0.17$ & $15.23\pm0.08$ & $\cdots$ & $13.52\pm0.07$ & $0.04\pm0.08$ & $3.01\pm0.11$ & 4, 8 \\
111008A & $0.12\pm0.04$ & 4.991 & $>16.04$ & $16.05\pm0.05$ & $15.71\pm0.09$ & $13.28\pm0.21$ & $0.44\pm0.06$ & $2.47^{+0.07}_{-0.09}$ & 3, 4 \\
111107A & $0.13\pm0.03$ & 2.893 & $15.87\pm0.20$ & $<14.70$ & $\cdots$ & $\cdots$ & $0.28\pm0.09$ & $2.37^{+0.18}_{-0.21}$ & 6, 9 \\
120119A & $1.02\pm0.11$ & 1.729 & $16.67\pm0.35$ & $15.95\pm0.25$ & $\cdots$ & $14.04\pm0.25$ & $0.22\pm0.10$ & $2.99^{+0.24}_{-0.18}$ & 3, 4 \\
120712A & $0.08\pm0.03$ & 4.172 & $>14.6$ & $\cdots$ & $\cdots$ & $\cdots$ & $0.47\pm0.09$ & $2.73^{+0.18}_{-0.23}$ & 9, 10 \\
120716A & $0.32\pm0.07$ & 2.487 & $16.48\pm0.45$ & $15.65\pm0.45$ & $\cdots$ & $13.91\pm0.32$ & $0.57\pm0.11$ & $2.84^{+0.21}_{-0.13}$ & 6, 11 \\
120815A & $0.19\pm0.04$ & 2.360 & $16.34\pm0.16$ & $15.29\pm0.05$ & $16.22\pm0.25$ & $13.47\pm0.06$ & $0.82\pm0.08$ & $2.38\pm0.09$ & 3, 4 \\
120909A & $0.09^{+0.04}_{-0.03}$ & 3.929 & $16.22\pm0.32$ & $15.20\pm0.18$ & $\cdots$ & $13.55\pm0.32$ & $0.58\pm0.10$ & $2.46^{+0.21}_{-0.18}$ & 6, 9, 11 \\
121024A & $0.26\pm0.07$ & 2.300 & $16.69\pm0.14$ & $15.82\pm0.05$ & $\cdots$ & $13.74\pm0.03$ & $0.42\pm0.07$ & $2.92^{+0.19}_{-0.14}$ & 4, 9, 12 \\
130408A & $0.22^{+0.04}_{-0.05}$ & 3.758 & $15.95\pm0.22$ & $15.52\pm0.11$ & $15.78\pm0.18$ & $12.87\pm0.16$ & $0.33\pm0.14$ & $2.83^{+0.14}_{-0.17}$ & 3, 4 \\
140311A  & $0.15^{+0.05}_{-0.04}$ & 4.955 & $>14.70$ & $15.78\pm0.15$ & $15.48\pm0.12$ & $13.28\pm0.14$ & $0.43\pm0.14$ & $2.66^{+0.17}_{-0.19}$ & 9, 10 \\
141028A & $0.13\pm0.04$ & 2.333 & $14.26\pm0.13$ & $14.29\pm0.10$ & $\cdots$ & $12.38\pm0.33$ & $0.41\pm0.12$ & $2.51^{+0.13}_{-0.14}$ & 6, 9 \\
141109A & $0.34\pm0.05$ & 2.994 & $>15.60$ & $15.54\pm0.04$ & $15.81\pm0.05$ & $13.18\pm0.06$ & $0.54\pm0.13$ & $2.74^{+0.17}_{-0.13}$ & 6, 9 \\
161023A & $0.09\pm0.03$ & 2.710 & $15.21\pm0.05$ & $14.80\pm0.03$ & $14.85\pm0.03$ & $\cdots$ & $0.37\pm0.11$ & $2.47^{+0.14}_{-0.11}$ & 6, 9 \\
180325A & $1.58^{+0.10}_{-0.12}$ & 2.249 & $16.12\pm0.48$ & $16.68\pm0.18$ & $\cdots$ & $13.13\pm0.14$ & $0.52\pm0.14$ & $4.58^{+0.37}_{-0.39}$ & 13 \\
\hline
\end{tabular}
\begin{minipage}{180mm}
1: \citep{heintz19}, 2: \citep{ardis}, 3: \citep[][and references therein]{zafar19d}, 4: \citep{zafar18a}, 5: \citep{postigo10}, 6: Zafar et al. (in prep), 7: \citep{postigo18}, 8: \citep{hartoog13}, 9: \citep{bolmer19}, 10: \citep{zafar18b}, 11: \citep{wiseman17}, 12: \citep{friis15}, 13: \citep{zafar18c}
\end{minipage}
\end{table*}

\subsection{Far-UV rise against dust-phase Si and Fe}
At $\lambda^{-1}>5.9\,\mu{\rm m}^{-1}$ of the extinction curve, $c_{4}^{\prime}$ only measures the far-UV non-linear rise and is not representative of the entire UV extinction. We observe no trend between $c_{4}^{\prime}$ with either $N$(Si)$_{\rm dust}$ or $N$(Fe)$_{\rm dust}$ (see bottom panels of Fig. \ref{comp}). A linear regression analysis of $c_{4}^{\prime}$ versus $N$(Si)$_{\rm dust}$ and $N$(Fe)$_{\rm dust}$ for our sample yields a Pearson's correlation coefficient $\rho=-0.07$ (significance $\alpha\approx16\%$) and $\rho=-0.12$ (significance $\alpha\approx36\%$), respectively. This is in contrast to the findings of \citet{haris16} who claimed with just a weak Pearson's correlation coefficient ($\rho=-0.32$) that silicon depletion correlates with the far-UV rise. As in \S\ref{rvdust}, we compared their $N$(Si)$_{\rm dust}$ of Galactic sightlines with $c_{4}^{\prime}$ values resulting in an even weaker Pearson's correlation coefficient $\rho=-0.17$ (significance $\alpha\approx47\%$), consistent with our results. In this parametrization of the extinction curve, $c_{4}^{\prime}$ represents the population of small grains. Previously, extinction curve models suggested that small silicate grains are responsible for the far-UV rise \citep{mathis77,draine84,weingartner01,clayton03}. However, \citet{mishra17} fitted the observed Galactic sightlines with a silicate-graphite dust model and reported that neither silicate nor carbonaceous dust correlates with the far-UV extinction. The far-UV extinction may therefore originate from a combined effect of both small carbon and silicate grains. A larger sample of more high quality measurements of extinctions and silicon and iron column densities will provide an even better understanding of the relation between the dust-phase refractory element abundances and ultraviolet extinction parameters.

We also note that both \citet{haris16} and \citet{mishra17} conducted their studies using Galactic sightlines usually exhibiting 2175\,\AA\ bump in their extinction curves, the bump being attributed to carbonaceous dust grains \citep{draine03}. Still, both studies have obtained discrepant results, likely due to the different methods used to estimate dust abundances (i.e., [Si/H]$_{\rm dust}=$ [Si/H]$_{\rm ISM}-$ [Si/H]$_{\rm gas}$. To derive silicon depletion, \citet{haris16} assumed the interstellar abundance of silicon to be solar. On the other hand, \citet{mishra17} considered three approaches to derive dust-phase abundances of Si: $i)$ considering two sets of interstellar abundances: the protosolar abundance of \citet{lodders03} and the B-star abundances of \citet{przybilla08}, $ii)$ using Kramers-Kronig Relation of \citet{purcell69}, and $iii)$ modelling the observed extinction curves using a silicate-graphite model. Additionally, they reported the results obtained from extinction curve modelling of the same Galactic sightlines.
	
In our study, out of the total 23 GRB afterglows only three GRB extinction curves exhibit the 2175\,\AA\ extinction bump (i.e. GRB\,070802 \citealt{ardis}; GRB\,080607 \citealt{perley11}; and GRB\,180325A \citealt{zafar18c}) Also we have used volatile element abundances to derive depletions. The difference in the methods of deriving depletions, types of extinction curves and hence environments and dust composition of the grains present in these environments may cause the discrepant results. We argue that our results are more reliable for high redshift studies where featureless extinction curves are more typically observed.

\subsection{Dust grain composition and sizes}
Silicon and iron are highly depleted elements in the early Universe. Their high abundance ratios are linked to high star formation rates \citep{tinsley79}. Nucleosynthesis yields of heavy elements also suggest a connection between iron and silicon abundance and Type II supernova (SN; \citealt{kobayashi06,kobayashi11}). Low and intermediate mass stars ($0.8<M_\star /M_\odot<8$) during their asymptotic giant branch phase \citep{gail10}, core-collapse SNe \citep{gall11} and grain growth in the ISM by selective mantle accretion \citep{draine09} are considered as viable mechanisms for dust formation. Particularly, core-collapse SNe are believed to produce most heavy elements in a very short timescale.

In the Galactic disk, the fractions of Si and Fe in the gas-phase decreases by a factor of ten between the warm and cold neutral medium \citep[CNM \& WNM;][]{savage96}. Silicon is more depleted in the CNM, while the iron is more in the WNM and is an important component of interstellar dust because of its comparable abundance to Mg and Si \citep{zhukovska16}. \citet{zhukovska16}, in their hydrodynamic simulation, attempted to explain the high depletion of Si in a Milky-Way like galaxy and find it to be a combination of accretion of Si on silicate grains in the CNM and efficient dust destruction by supernovae shocks in the diffuse gas. Most of the Si atoms have been locked up in silicate dust grains \citep{li05}. Based on condensation sequences, most of iron and magnesium are expected to condense in silicates and some Al, Na, and Ca might also be locked up in silicates \citep{tielens86}. The observed Fe depletion can be explained by the accretion of Fe onto dust grains \citep{dwek16}. A large fraction of interstellar iron is, for example, locked in silicate grains as demonstrated by the in-situ studies of dust grains \citep{westphal14,altobelli16}. The depleted Fe can be in form of metallic iron \citep{schalen65} with no extinction features, in silicate lattice \citep{ossenkopf92}, as pure iron and FeS to the silicates \citep{min07,jones13}, as iron oxides to the silicates \citep{draine13}, or in form of iron nanoparticles \citep{hensley17,gioannini17,bilalbegovic17}. The high depletion of both Si and Fe can also be explained by nanosized grains \citep{zhukovska18}.

Silicates are an important component in every dust grain model of the ISM \citep[e.g.,][]{mathis77,clayton03,siebenmorgen14,zhukovska16}. The models show that the efficient growth of silicate iron dust commences in the cold diffuse ISM \citep{zhukovska18}. Silicate dust grain formation has also been identified in SN ejecta \citep[e.g.,][]{haenecour13,sarangi15}. Silicate dust in the ISM is found to be entirely amorphous \citep[e.g.,][]{kemper04} and crystalline in addition to amorphous in protoplanetary disks \citep[e.g.,][]{sargent09}. \citet{decia16} used an observational sample of 70 damped Ly$\alpha$ absorbers towards quasars at intermediate redshifts and reported that after oxygen, which is ten times more abundant, Mg, Fe, and Si are other important dust constituents. The similar abundances of Mg, Si, and Fe (Mg:Si:Fe $\sim$1) in dust indicate that amorphous silicates such as pyroxenes (e.g., MgSiO$_3$ and FeSiO$_3$) and iron oxides (e.g., FeO and Fe$_2$O$_3$) are dominant grain species, while silicates such as olivine (e.g., Mg$_2$SiO$_4$ and Fe$_2$SiO$_4$) might be less important \citep{decia16}. Our results suggest that high Si and Fe depletion indicates the presence of more silicates in the ISM and variation in their sizes can alter the change in the shape of the individual featureless extinction curves.

On the other hand, carbonaceous dust grains are formed in the ISM when there is sufficient carbon available. When a large reservoir of metals is available, non-carbonaceous grains will form via grain growth in the ISM \citep{draine09}. \citet{decia16}, in their study of dust composition, suggested that the presence of carbon or the grain size distribution are responsible for extinction curves shapes, although carbon is not included in their analysis. \citet{decia16} reported that the main dust constituent is oxygen and up to 50\% of oxygen is locked in silicate grains and iron oxides. \citet{heintz19} and \citet{mishra17} found a correlation between $R_V$ and carbon (gas and dust phase) suggesting carbon grains as the possible carrier of the 2175\,\AA\ extinction bump. However, we find here that in addition to the abundance of carbon and the average grain size distribution, the amount of silicate grains can also alter the shape of the extinction curve. In our sample, 87\% (20 out of 23 GRBs) show featureless extinction curves, which implies that silicates play a dominant role to induce the featureless extinction curve shape when carbon grains do not dominate the dust content.

The interstellar dust size distributions extends down to nanometer sizes. Nanoparticles recovered meteorites could be a realistic model for interstellar grains \citep{mautner06}. Still the modelling of nanoparticles is challenging where not only the effects of environment on the particles but also their return effects (photo-electric heating of gas, formation of molecules etc.) should be taken into account \citep{jones16}. Smaller grains, like nano size silicate dust grains, may contribute to the far-UV extinction \citep{draine95} and could be carriers of observed extended red emission \citep{zubko99}. Although the exact sizes of these nano grains cannot be constrained by the far-UV extinction \citep{wang15}, and since they are suggested to account for only 5\% of total interstellar silicon, they have no significant effect on Si depletion \citep{li12}. Iron nanoparticles have also been proposed to partly contribute to the far-UV extinction \citep{hensley17}. Inclusion of iron nanoparticles within large silicate grains has been directly analysed  \citep{westphal14,hilchenbach16} and they are suggested to be a vital contributor to interstellar extinction \citep{kohler14}. Our results here suggest that the far-UV extinction could be a combined effect of small (or nano) dust grains from various species. Also high Si and Fe (since its large fraction is locked in silicates) depletion increases the fraction of large silicate grains in the environment and hence responsible for flatter extinction curves.

\section{Conclusions}
Understanding interstellar dust is important to study ISM and star-formation in galaxies. The attribution of the 2175\,\AA\ bump of the extinction curve to the carbonaceous dust is well studied for GRBs and quasar absorbers \citep{zafar12,ledoux15,ma18,heintz19a,heintz19}. We here compared extinction curve parameters to the Si and Fe dust-phase column densities to infer the composition of dust grains in the average GRB absorber population, showing typically featureless extinction curves. We analyzed a sample of 23 GRB afterglow ($0.6 \leq z \leq 5.0$) with available extinction curve parameters derived from X-ray$-$to$-$optical/NIR SED fitting and observed Si and Fe column densities. The majority of our sample (87\%) indeed do show featureless extinction curves. We then computed Si and Fe dust-phase column densities of these GRBs either estimating depletions using volatile element abundances (S and Zn) or using the dust-to-metals relation of \citet{zafar19d}. These Si and Fe dust-phase column densities are then compared against extinction, $R_V^{-1}$, and $c_4^{\prime}$ (far-UV component). 

We found a strong correlation between $R_V^{-1}$ and Si and Fe dust column densities with $>96$\% significance. Previously, \citet{decia16} found that carbonaceous grains and/or the dust grain size distribution are the main driver of the shape of the extinction curve. However, our results suggest that for featureless extinction curves, the shape of the extinction curve and dust grain size variation is linked to the amount of Si and Fe in dust-phase. High depletion of refractory elements (here Si and Fe) leads to an increase in the fraction of large silicate grains in the ISM and vice versa. A large fraction of iron is locked as inclusions in silicate dust grains. Therefore, this correlation indicates that silicates play an important role to induce the entire extinction at any wavelength. On the contrary, the far-UV extinction curve rise is suggested to be due to the presence of small silicate grains. However, we find no trend between far-UV parameter ($c_4^{\prime}$) and Si and Fe dust. Previously, \citet{mishra17} found no trend between carbon and silicon dust trend and far-UV rise for galactic sightlines. We, therefore, suggest that far-UV extinction could be a combined effect of small grains (probably nano-sized grains) from various species.



\bibliographystyle{aa}
\bibliography{silicate_dust.bib}{}

\bsp

\label{lastpage}
\end{document}